\documentclass [prl,twocolumn,superscriptaddress,showpacs,showkeys,preprintnumbers] {revtex4}
\usepackage[]{amssymb,amsmath,amsfonts}
\usepackage[]{graphicx}

\begin{document}
\title{Evidence of vectorial photoelectric effect on Copper}

\author{E. Pedersoli}
\affiliation{Dipartimento di Matematica e
Fisica, Universit\`a Cattolica, Via dei Musei 41, 25121 Brescia, Italy}

\author{F. Banfi}
\affiliation{Dipartimento di Matematica e
Fisica, Universit\`a Cattolica, Via dei Musei 41, 25121 Brescia, Italy}

\author{B. Ressel}
\affiliation{Sincrotrone Trieste S.p.A., Strada Statale 14, km
163.5, 34012 Basovizza (TS), Italy}

\author{S. Pagliara}
\author{C. Giannetti}
\author{G. Galimberti}
\affiliation{Dipartimento di Matematica e
Fisica, Universit\`a Cattolica, Via dei Musei 41, 25121 Brescia, Italy}

\author{S. Lidia}
\author{J. Corlett}
\affiliation{Lawrence Berkeley National Laboratory, One Cyclotron
Road, Berkeley, CA 94720 USA}

\author{G. Ferrini}
\author{F. Parmigiani}
\affiliation{Dipartimento di Matematica e
Fisica, Universit\`a Cattolica, Via dei Musei 41, 25121 Brescia, Italy}


\begin{abstract}

Quantum Efficiency (QE) measurements of single photon
photoemission from a Cu(111) single crystal and a Cu polycrystal
photocathodes, irradiated by 150~fs-6.28~eV laser pulses, are
reported over a broad range of incidence angle, both in $s$ and
$p$ polarizations. The maximum QE ($\simeq4\times10^{-4}$) for
polycrystalline Cu is obtained in $p$ polarization at an angle of
incidence $\theta=65^{\circ}$. We observe a QE enhancement in $p$
polarization which can not be explained in terms of optical
absorption, a phenomenon known as vectorial photoelectric effect.
Issues concerning surface roughness and symmetry considerations
are addressed. An explanation in terms of non local conductivity
tensor is proposed.
\end{abstract}

\pacs{79.60.Bm, 61.80.Ba, 41.85.Ar}

\keywords{Vectorial photoelectric effect, femtosecond
photoemission, copper photocathodes.}

\maketitle

The advent of the 4th generation free electron lasers (FEL)
sources \cite{Schoenlein, Neutze, Kapteyn} triggered several
important technical questions. A fundamental issue regards the
photocathode material for the laser-driven photoinjector devices,
to obtain short electron bunches with high charge density and low
emittance. Metal photocathodes are good candidates, having a high
reliability, long lifetime and a fast time response (1-10 fs).
However, two major drawbacks limit their usefulness, the small
quantum efficiency (QE) and the high work function ($\Phi$),
requiring light source in the ultraviolet (UV) for efficient
linear photoemission.

\begin{figure}[t!] \centering
\includegraphics[keepaspectratio,clip,width=9 cm]{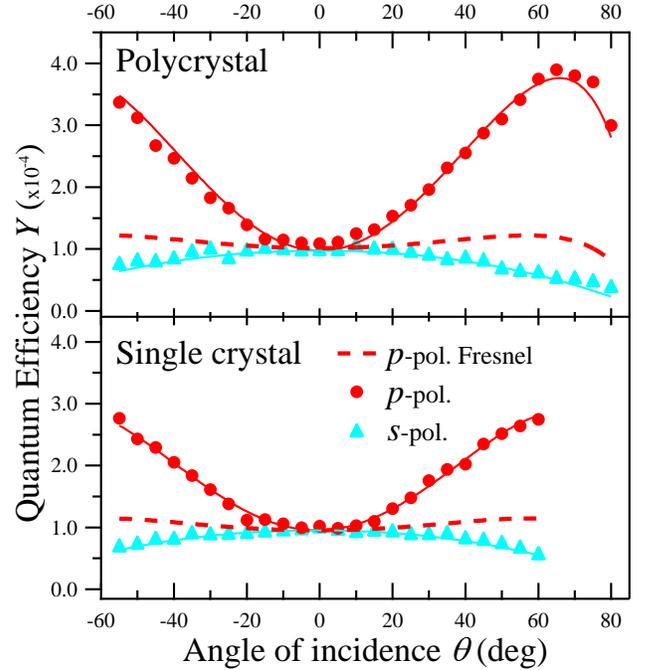}
\caption{Measurements of quantum efficiency dependence on the
angle of incidence $\theta$ for a Cu polycrystal and a Cu(111)
single crystal for $p$ (circles) and $s$ (triangles) polarized
light. Fits, based on Eq.~\ref{Efficiency}, are reported as solid
lines. The dashed lines are calculated taking into account Fresnel
absorption only.} \label{Measurement} \pagebreak
\end{figure}

In this Letter we study the experimental conditions to maximize
the QE of Cu photocathodes using UV short laser pulses from the
quadrupled output of an amplified Ti:Sapphire laser. The QE for
linear photoemission in the femtosecond regime is measured as a
function of the angle of incidence $\theta$ in the angular range
$-55^{\circ}\leq\theta\leq+80^{\circ}$, both in $s$ and $p$
polarizations. The maximum quantum efficiency
$Y\simeq4\times10^{-4}$, obtained with $p$ polarization at
$\theta=65^{\circ}$, is four times the value at normal incidence.

The QE dependence on angle of incidence and light polarization is
a long standing problem \cite{JWJ, Broudy1, Broudy, Girardeau,
Srinivasan} that largely remains to be understood. Our data are
well fitted by a phenomenological model \cite{Broudy} that keeps
into account only light absorption, without any explanation in
terms of microscopic quantum physics. A justification of the
phenomenological model based on the calculations of the
conductivity tensor for a jellium model is proposed.

The photoemission from a polycrystalline Cu sample and a Cu(111)
single crystal is investigated with 150~fs laser pulses with a
photon energy of 6.28~eV, obtained by two successive doubling
processes of the Ti:Sapphire fundamental frequency
($h\nu=1.57$~eV) in $\beta$-barium-borate (BBO) crystals. The
second doubling process is obtained out phase-matching in a thin
(200~$\mu$m) BBO crystal. The fourth harmonic is selected by
dispersing the doubling crystals output with a MgF$_2$ prism, with
minimal temporal and pulse front tilt distortions.

We do not use a more efficient third harmonic conversion to obtain
linear photoemission from Cu ($3h\nu=4.71$~eV, $\Phi=4.65~eV$ for
polycrystalline Cu \cite{Handbook2}) because of the onset of
multiphoton regime upon a work function increase due to sample
contamination. Moreover, an effective laser-induced oxide removal
and contaminants chemical-bond breaking obtained with UV short
laser pulses \cite{Afif1996, Beleznai1999} improves with shorter
wavelengths \cite{Afif1996}. Working with a 6.28~eV photon energy
should thus help to increase the duty time of machines based on Cu
photocathodes.

The quantum efficiency $Y$ is the ratio between the number of
photoemitted electrons, obtained from the photocurrent measured
from the sample with a Keithley 6485 Picoammeter, and the number
of incident photons, detected measuring on a Tektronix TDS3054B
digital oscilloscope the output of a Hamamatsu R928
photomultiplier tube. The measurements are performed with the Cu
photocathodes kept in a ultrahigh vacuum chamber with a base
pressure of $2\times10^{-10}$~mbar at room temperature. During the
total yield measurements, the photoemission spectrum is acquired
using a time of flight spectrometer in order to measure the sample
work function and monitor possible onset of sample contaminations
and space charge effects. The samples are cleaned by cycles of
Ar$^{+}$ sputtering followed by annealing at $500^{\circ}$C. This
procedure is continued until the proper value of the measured work
function (4.65~eV for the polycrystal and 4.94~eV for the single
crystal) is obtained. In these conditions a clear low energy
electron diffraction (LEED) pattern for the Cu(111) sample is
obtained.  The laser peak intensity on the target is
$I\simeq5\times10^5$~W/cm$^{2}$.

\begin{figure}[b]
\centering
\includegraphics[keepaspectratio,clip,width=9 cm]{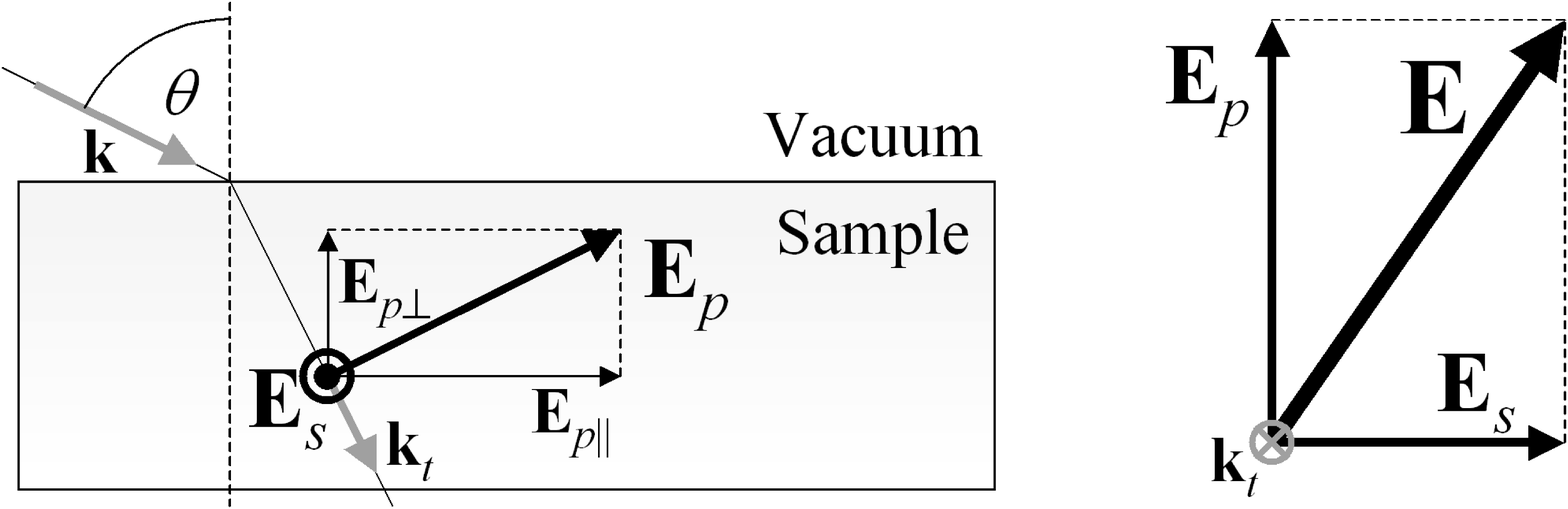}
\caption{Representation of incidence angle $\theta$, wave vectors
$k$ and $k_t$ for incident and transmitted light and field
components addressed in the text. A real index of refraction $n$
is assumed for the present figure.}\label{Vectors} \pagebreak
\end{figure}

The QE measured for both samples are reported in
Fig.~\ref{Measurement}. An enhancement of the QE is evident for
$p$ polarization as compared to what would be expected taking into
account only the electromagnetic absorption process. The maximum
QE do not occur at the pseudo-Brewster angle of incidence
$\theta_B=57^{\circ}$ \cite{commento}(see Fig.~\ref{Measurement}),
where there is maximum absorption, but is shifted by $\sim8^\circ$
toward the normal.

Our experimental results can be rationalized in the frame of a
phenomenological model proposed in Ref.~\onlinecite{Broudy}. The
electric field transmitted inside the sample can be written as
$\mathbf{E}=\mathbf{E}_{p}+\mathbf{E}_{s}=\mathbf{E_{\parallel}}+\mathbf{E_{\perp}}$,
where $\mathbf{E}_{p}$ and $\mathbf{E}_{s}$ are the $p$ and $s$
polarized field components respectively,
$\mathbf{E}_{\parallel}=\mathbf{E}_{p\parallel}+\mathbf{E}_s$ and
$\mathbf{E}_{\perp}=\mathbf{E}_{p\perp}$ are the components
parallel and perpendicular to the surface respectively. The
electric field vector components are defined in
Fig.~\ref{Vectors}. The QE, normalized with respect to its value
at normal incidence $Y(0)$, is:
\begin{equation}\label{Efficiency}
    \frac{Y(\theta)}{Y(0)}=\frac{\varepsilon_{\parallel}(\theta)}{\varepsilon_{\parallel}(0)}
    +r\frac{\varepsilon_{\perp}(\theta)}{\varepsilon_{\parallel}(0)},
\end{equation}
where $\varepsilon_{\perp}=\varepsilon_{p\perp}$ and
$\varepsilon_{\parallel}=\varepsilon_{p\parallel}+\varepsilon_{s}$
are the electromagnetic energies inside the sample due to the
fields components indicated by the suffixes. A value $r=1$ means that photoemission is
proportional to the absorbed intensity, whereas $r>1$ implies that
$\mathbf{E}_{\perp}$ is more efficient than
$\mathbf{E}_{\parallel}$ in producing photoelectrons.
Eq.~\ref{Efficiency} specialized for $p$ polarization is:
\begin{equation}\label{Efficiencyp}
   \frac{Y_p(\theta)}{Y_p(0)}=\frac{\varepsilon_{p\parallel}(\theta)}{\varepsilon_{p\parallel}(0)}
    +r\frac{\varepsilon_{p\perp}(\theta)}{\varepsilon_{p\parallel}(0)},
\end{equation}
whereas for $s$ polarization ($\mathbf{E}_{\perp}=0$):
\begin{equation}\label{Efficiencys}
    \frac{Y_s(\theta)}{Y_s(0)}=\frac{\varepsilon_s(\theta)}{\varepsilon_s(0)}.
\end{equation}
Once the electromagnetic energies
$\varepsilon_{p\parallel}(\theta)$, $\varepsilon_{p\perp}(\theta)$
and $\varepsilon_s(\theta)$ are calculated from classical
electrodynamics, assuming volume absorption as in
Refs.~\onlinecite{Broudy, Fan}, the parameter $r$ is obtained
fitting the experimental data for $p$ polarization with
Eq.~\ref{Efficiencyp}. The best fit values are $r=13$ for the
polycrystalline Cu  and $r=9$ for the Cu(111) single crystal (see
Fig.~\ref{Measurement}). The QE dependence expected on the basis
of Fresnel laws only, setting $r=1$, is also reported as a dashed
line in Fig.~\ref{Measurement}. The data for $s$ polarization are
in agreement with Eq.~\ref{Efficiencys}.

At the light of our data, it is important to investigate the
physical mechanisms that enhances the photoelectron yield due to $\mathbf{E}_{\perp}$ over
$\mathbf{E}_{\parallel}$.

The crystalline symmetry, important when dealing with polarization
dependent photoemission, play no role in the present
experiment. The photoemission process due to
$\mathbf{E}_{\perp}$ is about 10 times more effective than
$\mathbf{E}_{\parallel}$ both in the Cu(111) single crystal, where
symmetry considerations could apply, and in the polycrystalline Cu,
where any symmetry-related contribution is cancelled by the random
orientations of the single crystals domains composing the sample.

\begin{figure}[t]
\centering
\includegraphics[keepaspectratio,clip,width=9 cm]{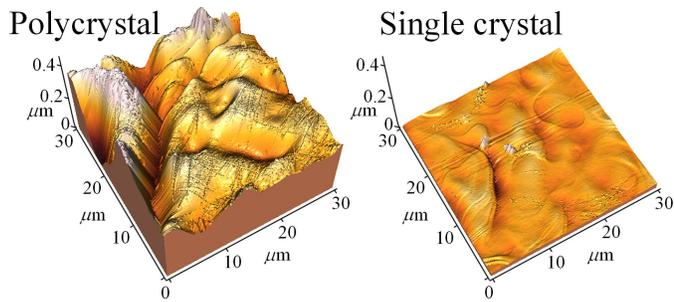}
\caption{Atomic Force Microscopy images of the two samples'
surfaces. Measured route mean squared roughness is 20~nm for the
Cu polycrystal and 2~nm for the Cu(111) single
crystal.}\label{Roughness} \pagebreak
\end{figure}

Photoemission enhancement due to surface roughness has been
recently investigated \cite{Banfi2003, Zawadzka2001,
Kupersztych2001}. In the present case surface roughness
enhancement can be ruled out. Several atomic force microscopy
(AFM) scans of the samples surface, with sizes ranging from
$1\times1~\mu \textrm{m}^2$ to $60\times60~\mu \textrm{m}^2$, give
values of the root mean squared roughness $h_{rms}\sim20$~nm for
the Cu polycrystal and $h_{rms}\sim2$~nm for the Cu(111) single
crystal, see Fig.~\ref{Roughness}. The observed vectorial
photoelectric effect is comparable on both samples, despite their
surface roughnesses differ by an order of magnitude. The
comparative study of the single crystal Cu and polycrystalline Cu
cathodes allows to clarify that our experiment is not dependent on
sample morphology.

Therefore, we seek for an explanation in terms of a more general
mechanism. Solutions of the Maxwell equations on an ideal
jellium-vacuum interface for an impinging plane electromagnetic
wave of frequency $\omega$, evidence an electromagnetic field
spatially varying on the length scale of $\sim1~\textrm{\AA}$ on
the jellium side \cite{FeibelmanPRB1975}. The spatially varying
electromagnetic field is due to the non local character of the
conductivity tensor. This is calculated using free-electron like
wave functions, so it does not depend on the symmetry of the
crystal. The matrix element entering the differential
cross-section for photoemission is composed of two terms. The
first is the usual electric dipole contribution, the second is due
to the rapidly varying electric field. The second term prevails
for $\omega<\omega_{p}$, where $\omega_{p}$ is the plasma
frequency, and leads to an enhancement of the photocurrent for the
electric field components perpendicular to the sample surface
\cite{FeibelmanPRL1975, Levinson}. In the present experiment,
$\hbar\omega=6.28$~eV and $\hbar\omega_{p}\sim19$~eV
\cite{Misell}. This mechanism explains an enhancement of the QE
for $p$ polarized incident radiation while not affecting the
results for $s$ polarized light. Furthermore, it does not depend
on surface roughness or a particular symmetry of the crystal. We
therefore propose it as the main microscopic mechanism to explain
our experimental evidences.

In this Letter quantum efficiency measurements on Cu
photocathodes, irradiated with 150~fs laser pulses at 6.28~eV, are
reported over a broad range of incident angles in both $s$ and $p$
polarizations. A QE enhancement is found for light with electric
field perpendicular to the sample's surface, showing a vectorial
photoelectric effect. The maximum value of quantum efficiency
$Y\simeq4\times10^{-4}$ is four times bigger than the QE at normal
incidence and is achieved with $p$ polarized light impinging on
the sample at an incidence angle of $\theta=65^{\circ}$.
Investigation of both a Cu(111) single crystal and a Cu
polycrystal allows us to rule out a microscopic processes based on
symmetry considerations and surface roughness to explain our data.
An explanation in terms of a rapidly varying effective field, due
to the non local character of the conductivity tensor, is
suggested.

This work was supported by the U.S. Department of Energy,
Office of Science, under Contract No. DE-AC03-76SF00098.

\bibliography{Bibliography}

\begin{thebibliography}{20}
\bibitem[{com()}]{commento}
\bibinfo{note}{An index of refraction $n=0.98+1.49i$ \cite{Handbook1} is
  assumed for Cu.}
\end{thebibliography}
\begin{widetext}  
Copyright (2005) American Institute of Physics. This article may be downloaded for personal use only. Any other use requires prior permission of the author and the American Institute of Physics.
The article appeared in Appl. Phys. Lett. 87, 081112 (2005); doi:10.1063/1.2031949 and may be found at http://link.aip.org/link/?apl/87/081112.
\end{widetext}

\end{document}